\documentclass[%
 reprint,
 amsmath,amssymb,
 aps,
]{revtex4-2}

\usepackage{graphicx}
\usepackage{color}
\usepackage{dcolumn}
\usepackage{bm}
\usepackage{comment}
\usepackage{url}

\begin{document}

\preprint{APS/123-QED}

\title{A diffuse scattering model of ultracold neutrons on wavy surfaces 
}

\thanks{Corresponding author: imajo@rcnp.osaka-u.ac.jp}%

\author{
S. Imajo,$^{1,*}$
H. Akatsuka,$^{2}$
K. Hatanaka,$^{1}$
T. Higuchi,$^{1}$
G. Ichikawa,$^{3,4}$
S. Kawasaki,$^{3}$
M. Kitaguchi,$^{2}$
R. Mammei,$^{5,6}$
R. Matsumiya,$^{1,5}$
K. Mishima,$^{3,4}$
R. Picker,$^{5,7}$
W. Schreyer,$^{5}$
and
H. M. Shimizu$^{2,3}$
}

\affiliation{%
$^{1}$Research Center for Nuclear Physics, Osaka University, Osaka 567-0047, Japan\\
$^{2}$Nagoya University, Nagoya 464-8601, Japan\\
$^{3}$KEK, Tsukuba 305-0801, Japan\\
$^{4}$J-PARC, Tokai 319-1195, Japan\\
$^{5}$TRIUMF, Vancouver, BC V6T 2A3, Canada\\
$^{6}$University of Winnipeg, Winnipeg, MB R3B 2E9, Canada\\
$^{7}$Simon Fraser University, Burnaby, BC V5A 1S6, Canada\looseness=-3
}%

\date{\today}

\begin{abstract}
Metal tubes plated with nickel-phosphorus are used in many fundamental physics experiments using ultracold neutrons (UCN) because of their ease of fabrication.
These tubes are usually polished to an average roughness of 25--$150\,\mathrm{nm}$.
However, there is no scattering model that accurately describes UCN scattering on such a rough guide surface with a mean-square roughness larger than $5 \,\mathrm{nm}$.
We, therefore, developed a scattering model for UCN in which scattering from random surface waviness with a size larger than the UCN wavelength is described by a microfacet Bidirectional Reflectance Distribution Function model (mf-BRDF model), and scattering from smaller structures by the Lambert's cosine law (Lambert model).
For the surface waviness, we used the statistical distribution of surface slope measured by an atomic force microscope on a sample piece of guide tube as input of the model.
This model was used to describe UCN transmission experiments conducted at the pulsed UCN source at J-PARC.
In these experiments, a UCN beam collimated to a divergence angle smaller than $\pm 6^{\circ}$ was directed into a guide tube with a mean-square roughness of $6.4 \,\mathrm{nm}$ to $17 \,\mathrm{nm}$ at an oblique angle, and the UCN transport performance and its time-of-flight distribution were measured while changing the angle of incidence.
The mf-BRDF model combined with the Lambert model with scattering probability $p_{L} = 0.039\pm0.003$ reproduced the experimental results well.
We have thus established a procedure to evaluate the characteristics of UCN guide tubes with a surface roughness of approximately $10 \,\mathrm{nm}$.

\end{abstract}

\maketitle


\section{\label{sec1}Introduction}
Ultracold neutrons (UCNs) are typically slower than $7\,\mathrm{m/s}$ and are totally reflected on the surface of materials with large scattering length densities, such as nickel (Ni) or beryllium (Be).
Because of this unique property, UCN can be transported like gas through guide tubes or stored in containers made of or coated by such materials.
Therefore, UCNs are a useful probe in high-precision measurements of the properties of the neutrons such as the neutron electric dipole moment (nEDM)~\cite{Ramsey1982,Golub1994,Harris1999,Pendlebury2015,Baker2014,Abel2020}, neutron lifetime~\cite{Serebrov2018,Ezhov2018,Materne2009}, quantization by gravity~\cite{Abele2009,Sedmik2019}, and others.

The TUCAN collaboration~\cite{Pierre2018,Ahmed2019,Martin2020,Higuchi2022} aims to search for the nEDM with a sensitivity of $10^{-27} \,\mathrm{e \cdot cm}$ (1 $\sigma$), which is an order of magnitude better than the current limit of $1.8 \times 10^{-26} \,\mathrm{e \cdot cm}$ (90\% C.L.)~\cite{Abel2020}, and is currently installing experimental equipment at TRIUMF.
The experimental plan is to produce UCN at a rate of $1.4 \times 10^{7} \,\mathrm{s^{-1}}$~\cite{Schreyer2020} by combining a spallation neutron source and a super-thermal converter of superfluid helium.
The UCN source will be connected to an nEDM cell by UCN guide tubes with a total length of $12 \,\mathrm{m}$, and the cell will be filled with polarized UCN to a density of about $200 \,\mathrm{cm^{-3}}$ at the beginning of the measurement~\cite{Sidhu2023}.

The transport efficiency of UCN from the source to the cell depends on the specularity of the inner surface of the guide tube.
Even when UCNs are totally reflected, the specular reflectance is always less than 100\% on realistic surfaces, and off-specular scattering typically occurs with a probability of 3--5\%, owing to the surface roughness~\cite{Golub1991,Steyerl1972,Sinha1988}.
The off-specular scattering causes UCNs to randomly diffuse, which significantly reduces the average downstream velocity of the UCN flow along the UCN guide in comparison to specular reflection.
As a result, more UCNs stay longer in the guide tube and the average number of times UCNs hit the wall increases.
This increase in the wall collisions reduces the transport efficiency of UCNs, since they can be lost due to acceleration from inelastic scattering or absorption by nuclei, represented by the measure of loss per bounce~\cite{Golub1991,Atchison2006}, or by entering gaps between guide tubes or defects in the coating.
Consequently, the surface roughness of the guide tube is one of the factors that diminish the statistical accuracy of nEDM experiments.
Since the design of the experimental setup and the experimental analysis are performed using UCN transport simulations that take into account off-specular scattering, the implementation of an accurate scattering model will improve the reliability of these simulations.

The guide tubes used in the TUCAN experiment are aluminum or stainless steel cylinders that are polished and coated with a nickel-phosphorus (NiP) alloy.
The NiP alloy has 11--13\% phosphorus by weight, the Fermi potential is approximately $210 \,\mathrm{neV}$~\cite{Schreyer2022, Pattie2017} and the corresponding critical neutron velocity is $6.3\,\mathrm{m/s}$.
Such NiP alloy plating with more than 10\% phosphorus has several advantages, such as being non-magnetic at room temperature, durable, chemically resistant, capable of being coated in any shape, easy to fabricate, and commercially available.
However, as they are metal tubes, their surfaces exhibit a high roughness with a root mean square (RMS) amplitude much greater than $5 \,\mathrm{nm}$ and random surface undulations on a micrometer scale.
There is currently no neutron scattering model capable of describing such a large surface undulation.

The model most often used to describe off-specular scattering in UCN transport simulations is the phenomenologically-constructed Lambert's cosine law, referred to as the ``Lambert model'' in this paper. 
This model is convenient for an approximate estimation due to the low computational resources required.
In this model, the luminous intensity to a given direction from a micro-area is distributed in proportion to the cosine of the emission angle from the surface normals.
It introduces the scattering probability per bounce as a constant $p_{L}$, independent of the angle of incidence, or momentum transfer to the surface, and surface structure.
The distribution of scattering directions is also independent of these factors.
Therefore, this model cannot accurately describe realistic scattering.

In a model introduced by Golub {\it et al}.~\cite{Golub1991}, UCN transport by metal pipes is described analytically by a rarefied gas flow theory.
In this model, the UCN transmission probability $W$ for a short or highly-specular cylindrical guide pipe is expressed by the following equation,
\begin{equation}
       W = \left(1 + \frac{3Zf}{8R(2-f)} \right)^{-1},
       \label{equ01}
\end{equation}
where $R$ is the tube radius, $Z$ is the tube length, and $f$ is the off-specular scattering probability per bounce.
This approach inherits the above-mentioned drawback by using the Lambert model to describe the off-specular scattering.
Furthermore, it has been reported that this approximation does not agree with the transport simulation calculated by the Lambert model for the highly specular guides that have been used in recent years, especially for short lengths of a few meters or less~\cite{Berceanu1973}.
It also assumes an isotropic UCN gas with a stable flow and is not applicable to the case of collimated UCN incidence or free diffusion of stored UCN from a small container.

In order to accurately describe UCN scattering, it is desirable to use a scattering model constructed with microscopic surface roughness information.
The micro-roughness model, referred to as the ``MR model'' in this paper, is a well-known model in UCN physics~\cite{Steyerl1972,Atchison2010}.
This model assumes that the surface roughness with RMS amplitude $b$ is isotropic and autocorrelated in the short range, and that the correlation function is described by a Gaussian with correlation length $w$ as its width.
In addition, the surface roughness is considered as a layer of thin potential, and the off-specular scattering is described by the interference of reflected waves at the potential boundaries.
Therefore, the scattering probability and angular distribution depend on the velocity of the UCN, the angle of incidence on the surface, and the values of $b$ and $w$.
In the Lambert model, the distribution of scattering directions of UCN is mostly in the direction of the surface normals, whereas in the MR model, the UCN has a lobe-like distribution around the direction of specular reflection.
In the case of a guide tube with a small surface correlation length $w$, the angular distribution of transmitted UCN is expected to be more strongly concentrated in the forward direction than in the Lambert model~\cite{Robson1976,Berceanu1973,Brown1975}.
However, for the reflected waves to be coherent, the surface roughness must be sufficiently small compared to the wavelength of UCNs.
Its magnitude is typically $b \leq 5\,\mathrm{nm}$.
Therefore, the calculations cannot deal with the large surface undulations of guide tubes such as those used in the TUCAN experiments.

In the description of neutron reflectometry, the model developed by Pynn~\cite{Pynn1992} is quite effective.
This model is constructed using the Distorted Wave Born Approximation (DWBA)~\cite{Messiah1958} and successfully unifies several prior scattering models, including the MR-model, while overcoming their drawbacks.
However, this model still requires approximations that assume spatially high-frequency surface roughness and a very thin roughness layer for specific calculations of scattering cross-sections.
For example, in the case of a NiP alloy containing 10\% phosphorus by weight, the layer thickness should have a standard deviation of the surface height irregularity, $\sigma_{z} \leq 3.5\,\mathrm{nm}$.
Therefore, even with this model, it is difficult to describe our UCN guide tubes.

\begin{figure*}[ht]
  \centering
  \includegraphics*[width=160mm]{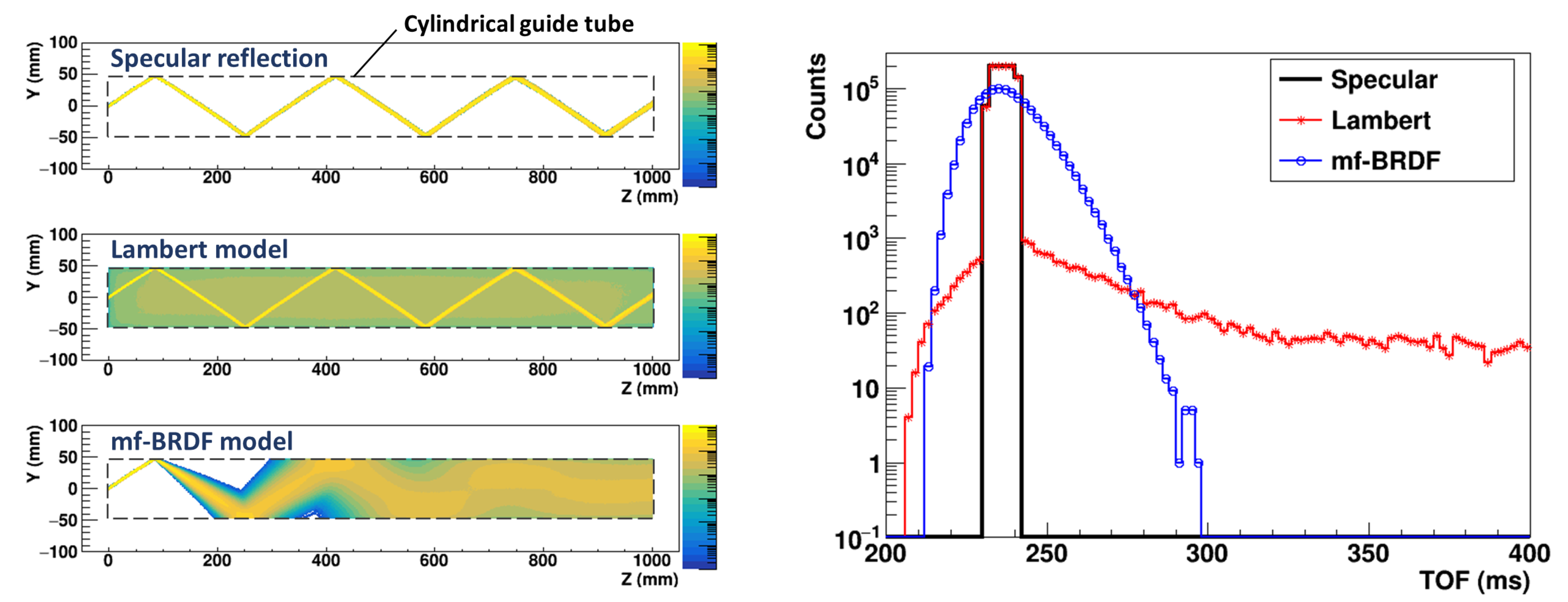}
  \caption{Differences in UCN trajectories (left panel) and TOF shapes (right panel) depending on the reflection models.
  They were calculated by our Monte Carlo simulation.
  In these calculations, $10^{6}$ parallel UCNs with a diameter of $5\,\mathrm{mm}$ and a velocity of $5\,\mathrm{m/s}$ are injected into a cylindrical guide tube with a length of $1\,\mathrm{m}$ at an upward angle of $30^{\circ}$. 
  The time distribution of the incident UCNs is a rectangular pulse with a width of $10\,\mathrm{ms}$.
  The value of $p_{L}$ of the Lambert model and that of $\alpha$ of the mf-BRDF model, which is explained in Section \ref{sec31}, are both set to 0.03. 
  The left panel shows the integration of the three-dimensional UCN trajectories projected onto the plane of the paper surface. 
  The right panel shows the TOF of the incident UCNs until they reach the downstream of the guide tube.  
  \label{fig_ex03}}
\end{figure*}

Scattering of rays based on geometrical optics by surface structures larger than the wavelength has been well studied in visible light, and the microfacet-based Bidirectional Reflectance Distribution Function model~\cite{Torrance1967,Cook1982,Walter2007,Trowbridge1975,Heitz2014,Heitz2016}, referred to as the ``mf-BRDF model'' in this paper, is the most basic and well known.
This model treats the surface roughness as an assembly of small specular surfaces (microfacets) like crumpled aluminum foil, and scatters the surface's normal directions by modeling the statistical distribution of the slopes of the microfacets.
Additionally, by also modeling the spatial distribution of the slopes of the surface, the surface luminance is calculated by taking into account the masking and shadowing of the optical axis due to the peaks and valleys of the surface.
Therefore, the scattering angle distribution depends on the angle of incidence.
Moreover, all incident rays are off-specularly scattered unlike the Lambert model and the MR model, where the off-specular scattering is determined probabilistically and the rays that do not contribute to these models retain specular reflections.
The distribution of scattering directions becomes lobe-like as in the MR model.
This model is typically used when the size of the surface undulations is sufficiently larger than the wavelength of the rays (in such cases, the undulations are referred to as surface waviness rather than roughness).
In the present day, this model is used for physically based rendering in computer graphics.
In such calculations, the non-specular reflections through complex processes such as the multiple reflections or the subsurface scattering are approximated by the Lambert model and added to this model to describe realistic metallic luster~\cite{He1991}.

It is expected that this model can be used to describe the surface of realistic guide tubes in the range where neutron optics is similar to visible light, especially on the wavelength scale of UCN and very cold neutrons (VCN), where the wavelength is sufficiently larger than the interatomic distances.
For example, the UCN beamline at PSI uses mainly nickel-molybdenum (NiMo) coated glass tubes with a surface roughness amplitude of less than $2\,\mathrm{nm}$, while stainless-steel tubes with a roughness of up to $10\,\mathrm{nm}$ and a storage vessel with a roughness of $400\,\mathrm{nm}$~\cite{Bison2020} constitute 20--25\% of the transport path.
In the ``ping-pong'' experiment performed at this UCN source~\cite{Bison2022}, where UCNs contained in a storage cell at beamport West-1 were transported to beamport South, the peak of the measured time spectrum was a little wider and delayed compared to a simulation by the Lambert model.
Furthermore, the contribution of the Lambert model to the overall transport system was larger than expected from NiMo guide tubes.
These phenomena may be explained in more detail by considering the mf-BRDF model.

To evaluate the consistency between these scattering models and real surfaces, it is necessary to measure the scattering angle profile of UCN.
The specularity of practical UCN guide tubes can be measured by using transmission of continuous or pulsed UCN~\cite{Daum2014,Pattie2017,Frei2010}, the transmission of pre-stored UCN~\cite{Blau2016,Bison2022}, or by attaching pinholes at the entrance and exit of the guide tube to allow UCN to temporarily reside in the guide tube~\cite{Pattie2017,Altarev2007}.
However, it is quite difficult to evaluate the scattering profile from the results of these experiments because of the use of not well-collimated UCN.
An experiment to compare the Lambert model with the MR model was performed by Atchison {\it et al.}~\cite{Atchison2010} by transporting a continuous stream of UCN with a limited angle of incidence through sample plates facing each other, and comparing the attenuation of the transport efficiency to a simulation.
However, the samples measured in this experiment were flat plates with a small surface roughness of $b = 1$--$3\,\mathrm{nm}$.
Practical guide tubes such as metal pipes have never been measured by this technique.

Therefore, we developed a method for evaluating a suitable scattering model for the surface of a practical guide tube with inspiration from measurements by Atchison {\it et al}.
In our method, a pulsed UCN beam with a divergence of $\pm 6^{\circ}$ or less and approximately half the diameter of a UCN guide tube is obliquely incident on the guide tube at angles ranging from $0^{\circ}$ to $30^{\circ}$, and both the transport efficiency and the time-of-flight (TOF) spectra of the transmitted UCN are measured.
As previously stated, the scattering probability and scattering angle profiles vary depending on the model.
For example, as shown in Fig.~\ref{fig_ex03}, in the Lambert model, the time information of the scattered UCN is close to that of a continuous flow, while the TOF pulse shape of the remaining UCN is preserved.
In contrast, in the mf-BRDF model, the dispersion of the UCN flight distance rapidly increases as the incident angle increases, due to repeated lobe-like diffusion around the specular direction.
As a result, the TOF pulse shape gradually widens and the peak shape becomes unsharp.
Therefore, this method allows us to find a scattering model that simultaneously satisfies the reduction in transport efficiency and the deformation of the TOF pulse shape by comparison with transport simulations.

Using this experimental approach, we confirm that a combination of the mf-BRDF model and the Lambert model explains well both the UCN transport efficiency and the deformation of the TOF pulse shape for our guide tube.
Noting the similarity between the optics of light and neutrons, we used these models to describe the scattering of UCN on the rough surface, assuming that geometrical optics holds in structures larger than the wavelength of a neutron.
In this paper, we discuss the experiments and the measurement results in Section \ref{sec2} and the details of the UCN scattering model and the results of the simulation analysis in Section \ref{sec3}.

\section{\label{sec2}Experiment}
The off-specular scattering causes attenuation of transmission and deformation of the TOF peak shape for the pulsed UCN flow in the guide tube~\cite{Daum2014}.
To measure this effect, an experiment was conducted using a neutron Doppler shifter~\cite{Mishima2014,Imajo2016}, which is a pulsed UCN source installed on the BL05 NOP beamline~\cite{Mishima2009,Arimoto2012} of the Materials and Life Science Experiment Facility (MLF) at the Japan Proton Accelerator Research Complex (J-PARC).
In the experiment, pulsed UCNs with velocities of $6$--$10\,\mathrm{m/s}$ were incident on a sample guide tube, and the TOF of the transmitted UCNs was measured while changing the angle of incidence on the guide tube between $0^{\circ}$, $10^{\circ}$, $15^{\circ}$, and $30^{\circ}$.
By increasing the inclination angle, the number of reflections increases from one to six, assuming specular reflection.
This increase in the number of reflections corresponds to an increase in the probability of having an off-specular scattering.
As a result, the UCN transport efficiency is attenuated as the inclination angle increases and the deformation of the pulse shape becomes stronger.
The validity of the scattering model is examined by comparing these observed changes with the simulations described in Section~\ref{sec3}.

\begin{figure*}[htbp]
  \centering
  \includegraphics*[width=160mm]{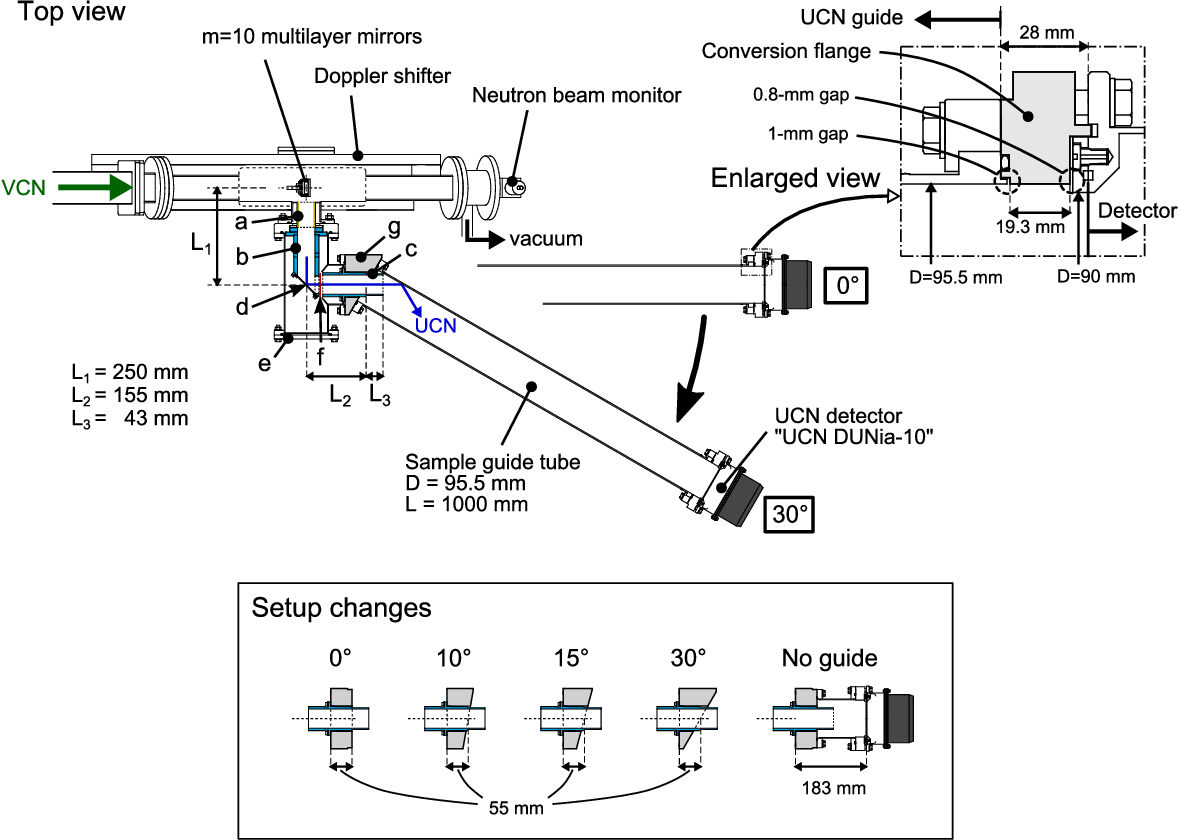}
  \caption{Setup of the UCN transmission experiment (Top view). The figure shows the setup with the sample guide tube attached at an angle of $30^{\circ}$ and the detector position at an angle of  $0^{\circ}$. The upper right inset shows an enlarged view of the connection between the guide tube and the UCN detector, and the lower inset shows the setup that is changed when the installation angle of the guide tube is changed. The symbols in the figure mean the following:
  (a) a square collimator with internal dimensions of $45\,\mathrm{mm} \times 45\,\mathrm{mm}$ and a length of $61\,\mathrm{mm}$, coated with Gd$_{2}$O$_{3}$ compound,
  (b) a square collimator with internal dimensions of $45\,\mathrm{mm} \times 45\,\mathrm{mm}$ made of polyethylene plate,
  (c) a cylindrical collimator with an internal diameter of $50\,\mathrm{mm}$ and a length of $155\,\mathrm{mm}$ made of polyethylene,
  (d) NiC mirror sputtered on a Si wafer,
  (e) vacuum chamber,
  (f) $10\,\mathrm{\mu m}$ thick copper film,
  (g) angled flange,
  ($L_{1}$) distance from the point of UCN production in the Doppler shifter to the center of the NiC mirror, which is $250\,\mathrm{mm}$,
  ($L_{2}$) distance from the center of the NiC mirror to the center of the sample guide tube inlet, which is $155\,\mathrm{mm}$,
  ($L_{3}$) distance from the center of the sample guide tube inlet to the exit of the cylindrical collimator, which is $43\,\mathrm{mm}$. 
  \label{fig1}
  }
\end{figure*}

\subsection{\label{sec21}Experimental setup}
Fig.~\ref{fig1} shows an overall view of the experimental setup.
The neutron Doppler shifter~\cite{Imajo2016} is shown in the upper left corner of the figure.
VCNs with velocities of $136\,\mathrm{m/s}$ incident from the left are reflected and decelerated by $m=10$ monochromatic multilayer mirrors~\cite{Hino2015} with a width of $30\,\mathrm{mm}$ rotating at a peripheral velocity of $68\,\mathrm{m/s}$.
The resulting UCN pulse generated every $120\,\mathrm{ms}$ is extracted.
The pulse duration of the UCNs immediately after generation estimated from a simulation is $2\,\mathrm{ms}$ full width at half maximum (FWHM), which corresponds to the spatial size of the UCN cloud being about half the width of the mirror.
The jitter of the mirror rotation period is $\pm 8\,\mathrm{\mu s}$, which is negligible in this measurement.

The UCNs pass through collimators (a), (b), and (c), and then are led inside the sample guide.
Because the UCN cloud produced in the Doppler shifter spreads very widely, the direction of the incident UCNs is almost entirely determined by the collimators.
The UCN beam was deflected with a NiC-sputtered 2-mm-thick silicon wafer~\cite{Hino2015} before entering the sample guide because of the limitation of the experimental space.
The surface roughness amplitude of the NiC is $0.3\,\mathrm{nm}$, so the reflection here can be considered sufficiently specular.
The optical distance along the central axis from the entrance of collimator (a) to the exit of the collimator (c) is $413\,\mathrm{mm}$, and the maximum divergence angle of UCN calculated geometrically from the collimator size is $\pm 6.6^{\circ}$.
Our UCN generation simulation showed that 99\% of incident UCNs have divergence angles within $-6.5^{\circ}$ to $1.7^{\circ}$ vertically and $\pm 5.4^{\circ}$ horizontally and the standard deviations are $1.8^{\circ}$ and $2.1^{\circ}$, respectively.

In this experiment, the sample guide tube was attached to the vacuum chamber via one of several angled flanges, which were inclined by $0^{\circ}$, $10^{\circ}$, $15^{\circ}$, and $30^{\circ}$ to the direction of the UCN flow, as shown in the lower inset of Fig.~\ref{fig1}.
These angled flanges were made of aluminum 5056 alloys with the Fermi potential of $55\,\mathrm{neV}$.
The flanges were cut at an angle while maintaining a constant distance of $L_{1}+L_{2}=405\,\mathrm{mm}$, as shown in Fig.~\ref{fig1}.
The machining accuracy of these angles was $\pm0.1^{\circ}$, which is sufficiently small compared to the divergence angle of the incident UCN.
Backscattered UCNs returning to the entrance of the sample guide tube were lost on the collimator (c) or the angled flange.

The sample guide tube was made of aluminum 6061-T6 alloy and was $1000\,\mathrm{mm}$ in length.
Its inner surface was polished by Irving Polishing \& Manufacturing, Inc.~\cite{Irving} to an average surface roughness of approximately 50--$100\,\mathrm{nm}$ and coated by Chem Processing Inc.~\cite{Chem} with high-phosphorus NiP, with a thickness of 5 microns.
The detector was connected to the sample guide tube via a conversion flange made of 304 stainless steel, as shown in the upper right inset of Fig.~\ref{fig1}.
The inner surface of the flange was polished but not mirror-like.
The optical TOF distance from the UCN source to the detector entrance along the central axis of the incident beam path was calculated by $L_{1}+L_{2}+1028\,\mathrm{mm} / \cos \theta_{g}$, where $\theta_{g}$ is the installation angle of the sample guide tube.
The mean free path of the UCN with a velocity of $8\,\mathrm{m/s}$ in the detector was estimated to be $13\,\mathrm{mm}$.
The total TOF distances, assuming fully specular reflections, were estimated to be $1446\,\mathrm{mm}$, $1462\,\mathrm{mm}$, $1483\,\mathrm{mm}$, and $1605\,\mathrm{mm}$ for the $0^{\circ}$, $10^{\circ}$, $15^{\circ}$, and $30^{\circ}$ installation angles, respectively.
Considering the FWHM pulse width of $2\,\mathrm{ms}$, an $8\,\mathrm{m/s}$ UCN cloud with a width of $16\,\mathrm{mm}$ is produced within the width of the $30\,\mathrm{mm}$ multilayer mirrors.
Therefore, if we consider the actual starting point of the TOF distance to be the center of mass of the UCN cloud at their production, then the uncertainty of the TOF origin is less than 0.6\% of the TOF distance and is considered negligible for this measurement.

For UCN detection, a $^{3}$He proportional counter (DUNia-10 produced by A. V. Strelkov) was used.
The inner diameter of the sample guide tube and the conversion flange is $95.5\,\mathrm{mm}$, while the diameter of the aluminum window of the detector with a thickness of $100\,\mathrm{\mu m}$ is $90\,\mathrm{mm}$.
To shield the detector from background neutrons originating from the cold neutron beam leaking from the shield upstream of the Doppler shifter, the entire setup illustrated in Fig.~\ref{fig1} is covered with $5\,\mathrm{mm}$ thick B$_{4}$C rubber sheets.
The pulse height of the neutron detection signal and the time from the start of the measurement were recorded by an ADC/TDC system (Nikiglass A3400) with a time resolution of $1\,\mathrm{\mu s}$.
A $10\,\mathrm{\mu m}$ thick copper film (f) was placed between collimators (b) and (c) to prevent a frame overlap of the Doppler shifter's $120\,\mathrm{ms}$ pulses period by reflecting UCNs with velocities slower than $5.7\,\mathrm{m/s}$.

The entire apparatus was evacuated using a dry roughing pump through the vacuum outlet of the Doppler shifter itself, and measurements were performed in a vacuum-pressure range of 10--$30\,\mathrm{Pa}$.
The experiment was performed with the MLF proton beam power of $730\,\mathrm{kW}$.
The UCN production was normalized by the counts of a beam monitor located in the VCN beam path downstream of the Doppler shifter.

\subsection{\label{sec22}Measurements and results}
The TOF of the transmitted UCNs was measured by changing the installation angle of the sample guide tube to $0^{\circ}$, $10^{\circ}$, $15^{\circ}$, and $30^{\circ}$.
Incident neutrons were measured using the setup labeled ``No guide'' in the lower inset of Fig.~\ref{fig1}.
The count rate of incident UCNs was $1.53 \pm 0.01\,\mathrm{neutrons/s}$ at $730\,\mathrm{kW}$, and the count rate of UCNs detected through the sample guide tube was $1.33\,\mathrm{neutrons/s}$ at $0^{\circ}$ and $0.87\,\mathrm{neutrons/s}$ at $30^{\circ}$, decreasing with the increasing installation angle of the sample guide tube.
Incident UCNs were measured for 3.7 hours, and UCNs through the sample guide tube were measured for about 1 hour at each angle.
The cold neutron background was measured every time the setup was changed with the Doppler shifter stopped and the multilayer mirrors placed outside of the incident beam path.
The background count rate was $0.03 \,\mathrm{neutrons/s}$.
Bursts of fast neutrons produced at the moment of proton injection into the MLF target have a much higher peak flux than UCN, and thus introduce large statistical errors when background data are subtracted.
Therefore, data in the $-10\,\mathrm{\mu s}$ to $100\,\mathrm{\mu s}$ range of proton incidence were excluded from all analyses.

\begin{figure}[htbp]
  \centering
  \includegraphics*[width=80mm]{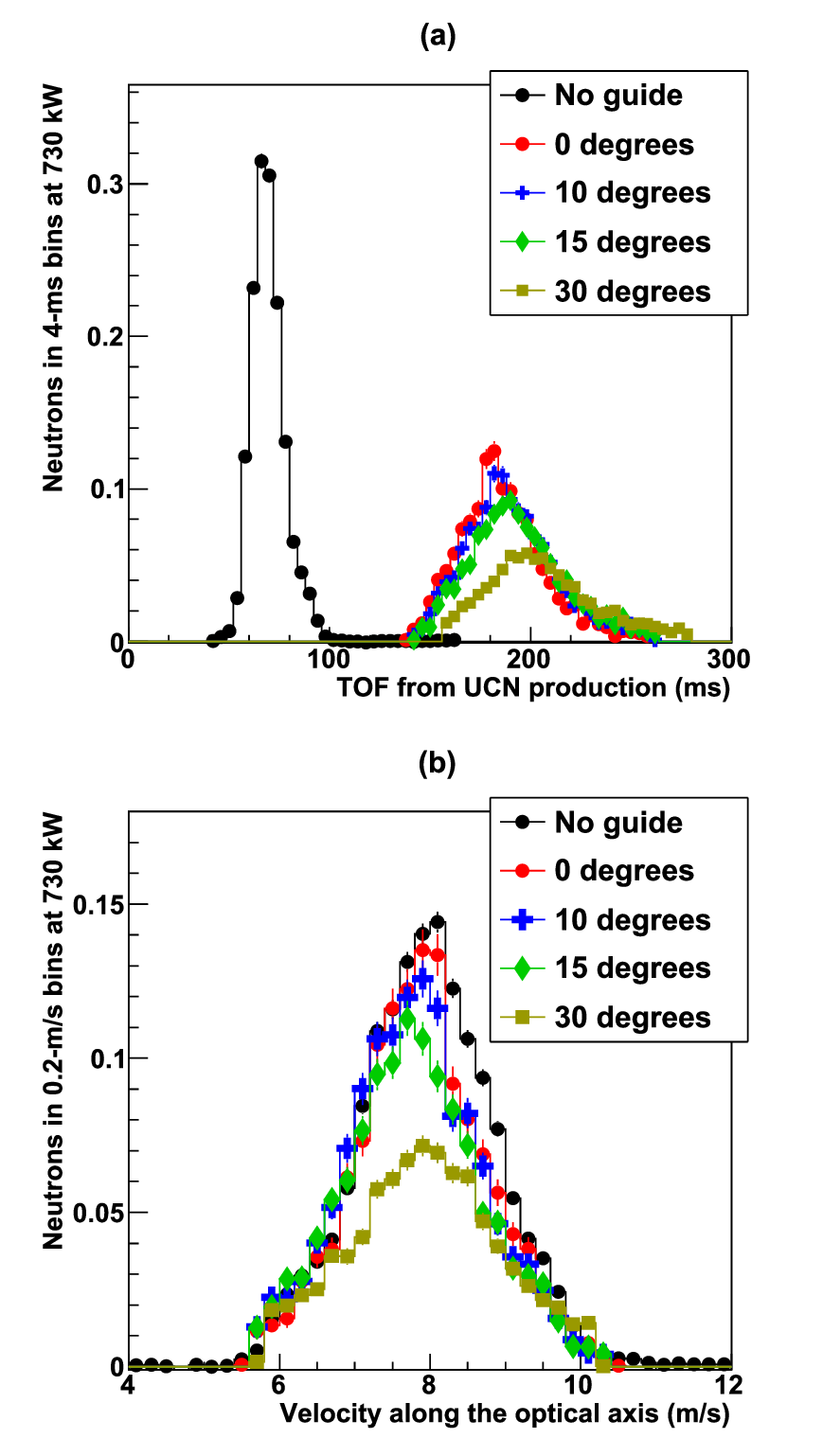}
  \caption{(a) TOF spectra of transmitted UCN measured without the sample guide tube, and with the tube at different installation angles. (b) Velocity distributions converted from (a) using TOF distances.\label{fig3}}
\end{figure}

The measured TOF and the velocity distribution obtained from the TOF and the flight distance along the optical axis are shown in Fig.~\ref{fig3}.
Since both TOF values of the UCNs passing through the sample guide tube and the total width of the TOF pulses exceeded the $120~\mathrm{ms}$ pulse period, the $120~\mathrm{ms}$ time-window used for analysis had to be artificially introduced in the continuous time information.
Thus, the time window of Fig.~\ref{fig3} (a) was chosen such that the center of the velocity distribution is $8~\mathrm{m/s}$.
The velocity distribution has a peak at $8\,\mathrm{m/s}$ at all installation angles, but at $30^{\circ}$ the peak becomes wider and unsharp.

\begin{figure}[htbp]
  \centering
  \includegraphics*[width=80mm]{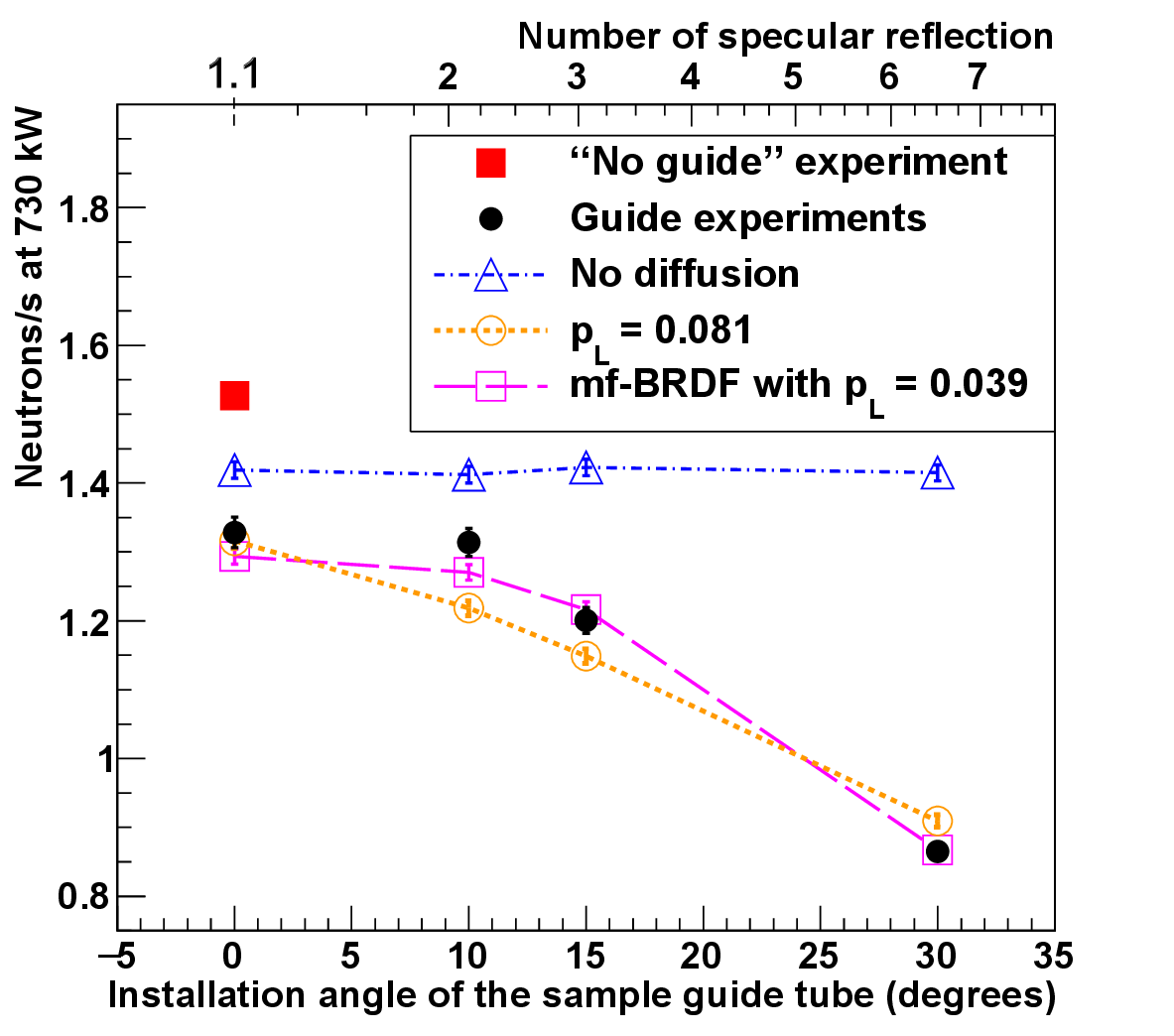}
  \caption{(Color online) Measured rate of transmitted neutrons with no guide (a red square) and with sample guide installed at different angles (black dots), compared to simulations with no diffuse reflection (blue triangles), pure Lambert scattering with probability $p_{L} = 0.081$ (orange circles), and mf-BRDF scattering combined with Lambert scattering with probability $p_{L} = 0.039$ (pink squares). Due to gravity and incidence angle, UCN has to undergo between 1 and 7 specular reflections to reach the detector, reducing the transmission compared to the no-diffusion simulation. \label{fig4}}
\end{figure}

Fig.~\ref{fig4} shows the transport efficiency obtained by integrating the spectra in Fig.~\ref{fig3}.
Also shown are the results of the transport calculation without surface roughness and the best-fit results using the Lambert model and the mf-BRDF model simulations, which will be discussed later in Section~\ref{sec3}.
The transmittance decreases with increasing installation angle of the sample guide tube as shown in Fig.~\ref{fig4}.
The UCN transmittance at $0^{\circ}$ and $30^{\circ}$ is decreased from the incident UCNs by factors of 0.87 and 0.57, respectively.
If no surface roughness existed and all reflections were specular, the average number of reflections at $0^{\circ}$ and $30^{\circ}$ would be 1.1 and 6.5, respectively.
As a result, in this scenario, the average reflectivity per reflection, considering transmittance as a power function of the number of reflections and average reflectivity as a variable, would be 0.92.
In comparison, the energy-independent loss probability $\bar{\mu}$ per reflection for this guide is as small as $(2.0$--$3.9) \times 10^{-4}$ (guide \#1 in Table 3 of \cite{Schreyer2022}).
Thus, this attenuation can only be explained by considering surface roughness.

\section{\label{sec3}Analysis with simulations}
The off-specular scattering of UCNs in a cylindrical tube and the resulting effects of multiple reflections were evaluated by a simulation.
We used the Lambert model and the mf-BRDF model as UCN scattering models to explain our experimental results.
In this section, we describe the details of our simulation analysis and the results.
The simulations are three-dimensional (3D) Monte Carlo particle transport calculations based on ray tracing methods including gravity.
\begin{figure*}[htbp]
  \centering
  \includegraphics*[width=130mm]{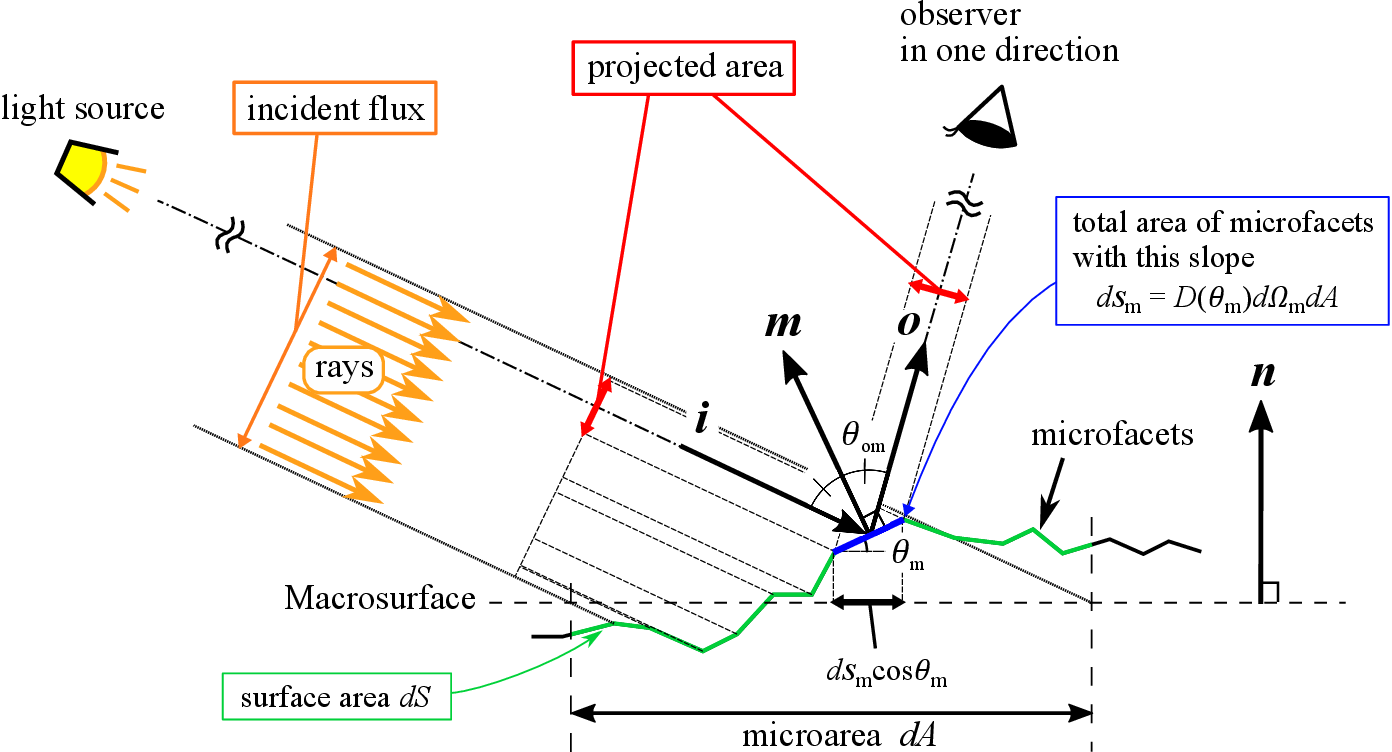}
  \caption{The off-specular scattering of incident rays due to surface roughness in the mf-BRDF model~\cite{Heitz2014, Walter2007}.\label{fig_ex1}}
\end{figure*}

\subsection{\label{sec31} Microfacet-based BRDF model }
In this model, as shown in Fig.~\ref{fig_ex1}, we consider the case where parallel rays having a unit vector $\bm{i}$ as their direction vector are incident on a surface of an object.
The surface consists of microfacets having randomly oriented unit normal vectors $\bm{m}$.
In this figure, an average height plane of microfacets, having a unit normal vector $\bm{n}$, is described as ``Macrosurface'' and a dashed line.
We assume that the micro-area $dA$ on the macrosurface is the area that is being irradiated.
A direction of $\bm{m}$ is then probabilistically selected, and the unit reflection vector $\bm{o}$ is given by $\bm{o} = \bm{i} - 2(\bm{i} \cdot \bm{m})\bm{m}$ according to the specular reflection.
This means that the direction of vector $\bm{o}-\bm{i}$ always coincides with vector $\bm{m}$.
Subsequently, the quantity of rays that an observer in the direction of $\bm{o}$ obtains via a group of microfacets with $\bm{m}$ as their normal is evaluated.
This quantity is proportional to the projected area of the microfacets that can be optically seen by the observer, as indicated by the ``projected area'' in Fig.~\ref{fig_ex1}.
Eventually, in the case of 100\% reflectivity, the scattering probability is expressed as the ratio of the geometric cross-section of rays arriving at the observer to that of the incident rays.

To estimate the distribution of scattering directions of a single particle incident on the micro-area, we used the following equation based on Ref.~\cite{Walter2007},
\begin{eqnarray}
       \Phi_{o} &=& \frac{1}{|\cos \theta_{i}|dA}\int \frac{|\cos \theta_{om}|}{|\cos \theta_{m}|} F(\theta_{i}, \theta_{m}) \nonumber \\
       &&\times \left\{ D(\theta_{m} )|\cos \theta_{m}| \right\} G(\theta_{i}, \theta_{o}, \theta_{m}) d \Omega_m,  \label{equ02}
\end{eqnarray}
where $\theta_{i}$, $\theta_{m}$, $\theta_{o}$ are the polar angle of the ray source direction, microfacet normal, and outgoing direction of the particle, $\theta_{om}$ is the angle between the vectors $\bm{o}$ and $\bm{m}$, $\Phi_{o}$ is the total reflected flux from the micro-area, $1/(|\cos \theta_{i}|dA)$ is the incident flux of the particle, and $F$ is the Fresnel reflectivity of the microfacet.
$D$ is the normal distribution function (hereafter NDF) of the microfacet, and $G$ is the shadowing-masking function of the surface.
Several models exist for $D$ and $G$, and in this paper, we use the Beckmann NDF.
In this case, $D$ is expressed as follows,
\begin{equation}
      D(\theta_{m} ) = \frac{\chi^{+}(\cos \theta_{m})}{\pi \alpha^{2} \cos^{4} \theta_{m}} \exp \left(-\frac{\tan^{2} \theta_{m}}{\alpha^{2}} \right),  \label{equ03}\\
\end{equation}
where $\alpha$ is the width of the NDF (diffusion width), and $\chi^{+} (x)$ is the Heaviside step function.
$G$ is the shadowing-masking function of the surface, and using the shadowing-masking function calculated with Smith's model for the Beckmann NDF given in Eq.~\ref{equ03}, we obtain
\begin{eqnarray}
      G(\theta_{i}, \theta_{o}, \theta_{m}) &=& \frac{2 \chi^{+} \left( \frac{\cos \theta_{im}}{\cos \theta_{i}} \right)}{1 + \mathrm{erf}( a_{i}) + \frac{1}{a_{i} \sqrt{\pi}} e^{-a_{i}^{2}}} \nonumber \\
      &&\times \frac{2 \chi^{+} \left( \frac{\cos \theta_{om}}{\cos \theta_{o}} \right)}{1 + \mathrm{erf}( a_{o}) + \frac{1}{a_{o} \sqrt{\pi}} e^{-a_{o}^{2}}},  \label{equ04}\\
      && a_{i} = \frac{1}{\alpha \tan \theta_{i}}, \hspace{2mm} a_{o} = \frac{1}{\alpha \tan \theta_{o}}, \label{equ_ex01}
\end{eqnarray}
where $\theta_{im}$ is the angle between the incident ray axis and the vector $\bm{m}$, thus $\theta_{im} = \theta_{om}$, and $\mathrm{erf}(x)$ is the error function.
The shadowing-masking function is the probability that the optical axis is not interrupted by surface undulations in both directions $\bm{i}$ and $\bm{o}$, and rapidly approaches zero near $90^{\circ}$ of $\theta_{i}$ and $\theta_{o}$.
This term suppresses the occurrence of a reflection axis that traces the macrosurface.

In the calculation of the scattering angle $(\theta_{o},\phi_{o})$, the azimuth angle of vector $\bm{i}$ was constrained to $\phi_{i}=0$.
In addition, because the NDF $D$ is expressed more simply as a function of $\theta_{m}$ than $\theta_{o}$, we generated random numbers of $(\theta_{m},\phi_{m})$ pairs using the inverse function method according to the section 5.2 in Ref~\cite{Walter2007}.
In this case, the direction distribution of the 3D vector $\bm{m}$ represents axial symmetry, whereas the direction of the vector $\bm{o}$ is concentrated around the angle $(\theta_{i},0)$ since $\bm{i}$ and $\bm{o}$ satisfy the relation of specular reflection with respect to $\bm{m}$.
Therefore, we computed Eq.~\ref{equ02} and obtained the $(\theta_{o},\phi_{o})$ distribution using the Monte Carlo integration by taking into account the following Jacobian,
\begin{equation}
    d\Omega_{m}=\left\| \frac{d\Omega_{m}}{d\Omega_{o}} \right\|d\Omega_{o} = \frac{1}{4 \cos \theta_{om}}d\Omega_{o}.       \label{equ09}
\end{equation}

\begin{figure}[htbp]
  \centering
  \includegraphics*[width=70mm]{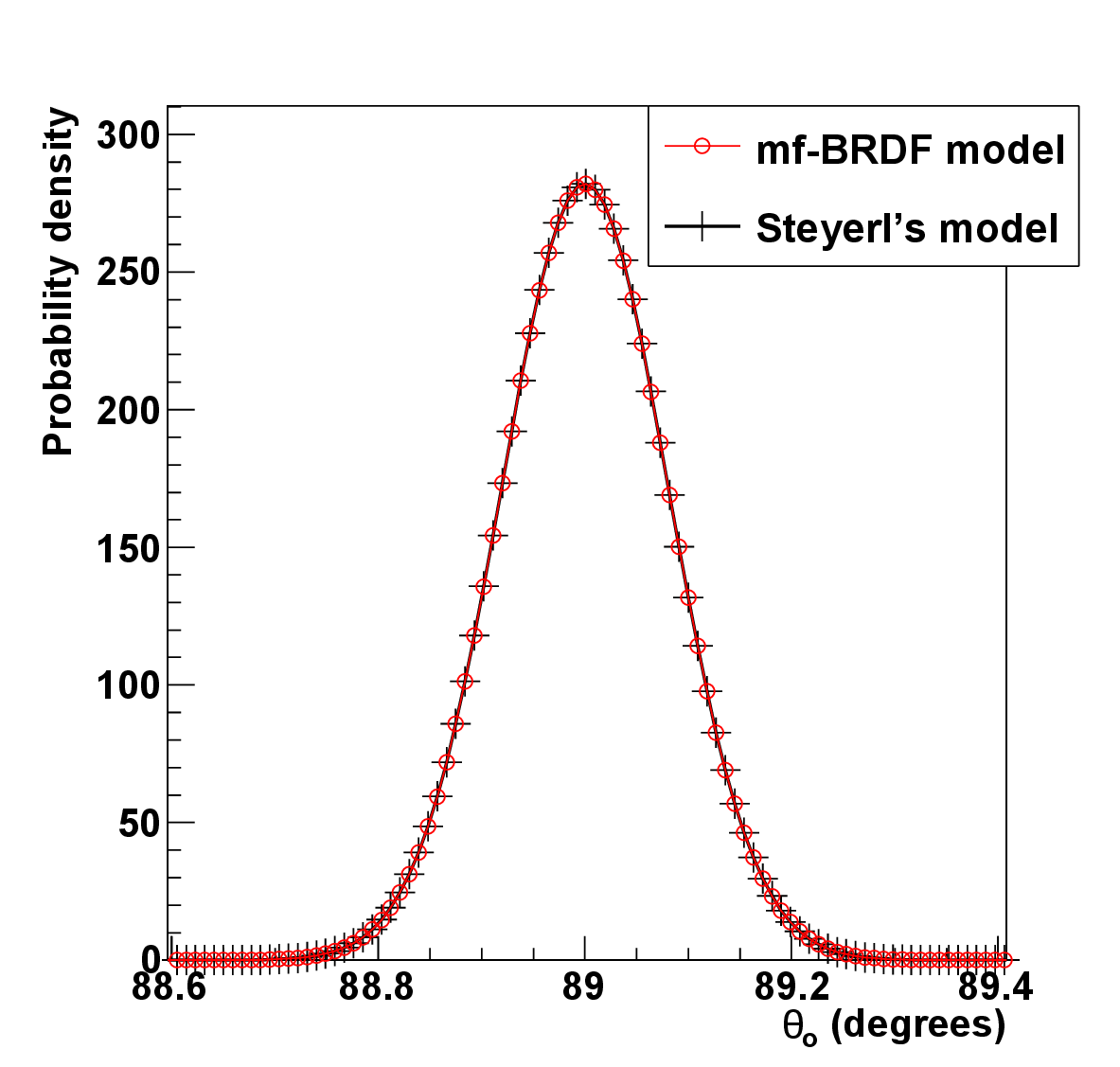}
  \caption{Comparison between the mf-BRDF model and a macroscopic waviness model described by Steyerl (Eq. (26) in Ref.~\cite{Steyerl1972}). They were calculated for $\alpha = 0.001$, $\theta_{i} = 89^{\circ}$, $\phi_{i} = 180^{\circ}$, and $\phi_{o} = 0^{\circ}$. In the mf-BRDF model calculation, $F$ was set to 1, but $G$ was calculated exactly. The curve for the mf-BRDF model is normalized to have an integral value of 1, and the curve for the Steyerl's model is normalized to the same value. \label{fig_ex04}}
\end{figure}

The Beckmann NDF is equivalent to the probability distribution of the two-dimensional (2D) slope of a microfacet described by a 2D normal distribution with mean zero and isotropic.
$\alpha$ is essentially the same as the RMS of the slope and Eq.~\ref{equ03} satisfies the following equation.
\begin{equation}
        \int D (\theta_{m} ) |\cos \theta_{m}| d \Omega_m = 1.       \label{equ05}
\end{equation}
From this relationship, this distribution can be regarded as calculating the percentage of the projection area of numerous microfacets with slope $\tan \theta_{m}$ in the micro-area shown in Fig.~\ref{fig_ex1}.
Letting $\alpha = \sqrt{2}b/w$ and $F=G=1$, the reflected amount $\Phi_{o}dA$ in Eq.~\ref{equ02} agrees with the macroscopic waviness described by Steyerl (Eq. (26) in Ref.~\cite{Steyerl1972}) for very shallow incident angles to the surface and for the approximations $\alpha \ll \cos \theta_{i}$ and $\cos \theta_{i} \sim \cos \theta_{o}$ (see Fig.~\ref{fig_ex04}).
Also, $\alpha=0$ gives a perfect specular reflection, $\alpha=0.01$ to 0.1 visually renders a buffed metal surface, and $\alpha \sim 1$ is close to the Lambert model with $p_{L} = 1$ (Fig. 2.4 in~\cite{Dupuy2015}).

\begin{figure}[!htbp]
  \centering
  \includegraphics*[width=80mm]{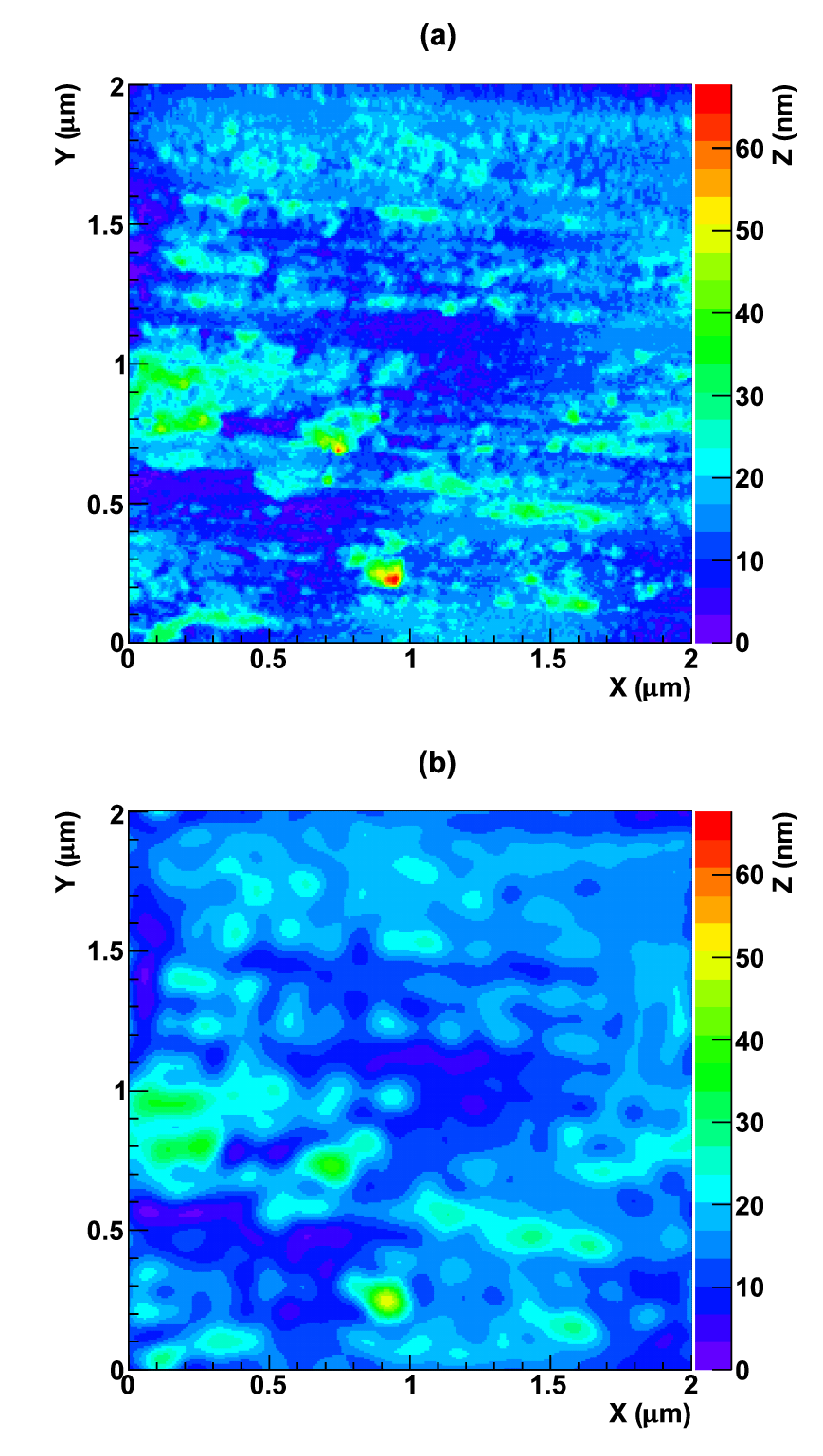}
  \caption{(a) Image of the surface of a guide tube sample measured by AFM. (b) Image obtained by frequency filtering to remove structures with a period of less than $128\,\mathrm{nm}$ from (a). \label{fig5}}
\end{figure}

In this mf-BRDF model, scattering is determined by $\alpha$ only.
The off-specular scattering can therefore be modeled by an $\alpha$ estimated by the normal vector distribution of the rough surface measured with an atomic force microscope (AFM) or similar instrument at the UCN wavelength scale.
To evaluate $\alpha$, we cut out a piece of a guide tube fabricated by the same process as the sample guide tube and observed its surface using AFM.
We observed the sample in two measurement ranges, measuring several different locations on the surface in each setting.
The measurement ranges were $2\,\mathrm{\mu m} \times 2\,\mathrm{\mu m}$ and $10\,\mathrm{\mu m} \times 10\,\mathrm{\mu m}$, with a resolution of $7.8\,\mathrm{nm}$ and $39\,\mathrm{nm}$ ($256 \times 256$ pixels), and the number of measured locations was 8 and 2, respectively.
Fig.~\ref{fig5} (a) shows one of the results measured in a $2\,\mathrm{\mu m} \times 2\,\mathrm{\mu m}$ view.
The average RMS of the surface roughness amplitude of 8 measurements with this scale was $b = 6.4 \,\mathrm{nm}$, while, that of 2 measurements at $10\,\mathrm{\mu m}$ scale was $b = 17 \,\mathrm{nm}$. 
As this figure shows, there are many bumps of 20--$30\,\mathrm{nm}$ in height on the guide surface.

The mf-BRDF model is a scattering model based on classical calculations of specular reflection. Furthermore, within this paper, we do not take into account the dispersion of the scattering direction due to quantum mechanical interference of a neutron scattered by nano-scale irregularities, and instead, we replace it with the Lambert model.
In this context, we perceive the smallest spot size, where a neutron can stochastically arrive, as the minimal interference region—in other words, the smallest area where we can disregard internal structures within this modeling scheme.
This size typically amounts to $\lambda / \cos \theta_{i}$, corresponding to a neutron's momentum perpendicular to the surface.
We chose this size as a cutoff wavelength $\lambda_{c}$ for the Fourier series expansion of surface waviness.
Then, we removed the roughness at shorter wavelengths using Fourier filter processing.
Fig.~\ref{fig5}(b) shows an example of filtering with a cutoff wavelength of $\lambda_{c} = 128,\mathrm{nm}$.

\begin{figure}[htbp]
  \centering
  \includegraphics*[width=70mm]{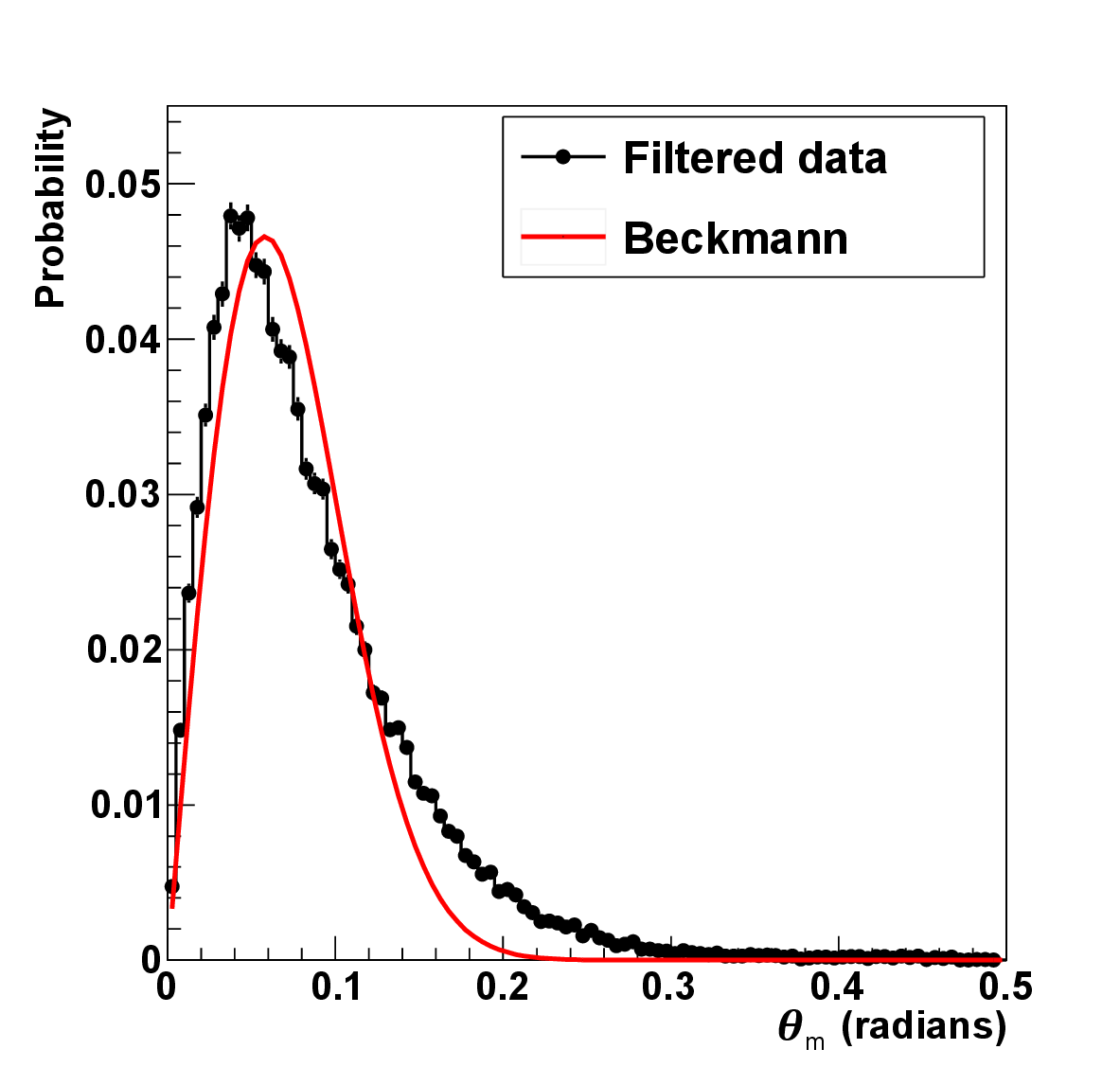}
  \caption{The polar angle distribution of the normal vector created from Fig.~\ref{fig5} (b) and the fit curve using Eq.~\ref{equ03}.
\label{fig6}}
\end{figure}

After processing the AFM image by a cutoff, normal vectors of each data point on the processed image were calculated, and a distribution of polar angles $\theta_{m}$ was created.
Specifically, the data grid was divided into a mesh pattern taking care to maintain surface continuity.
This was achieved by connecting the measurement points on the $x$--$y$ plane in Fig.~\ref{fig5} with lines along the $x$-axis and the $y$-axis, as well as the $45^{\circ}$ diagonals from the $x$-axis to prevent the creation of saddle points.
It should be noted that another division pattern exists, using diagonals of $135^{\circ}$ from the $x$-axis, and these cannot be unambiguously determined.
Subsequently, for a point of interest, we calculated the area vectors of the six surrounding triangles in 3D space that share this point.
The summed and normalized vector of these six was then adopted as the normal vector $\bm{m}$ of the point.
Also, we confirmed that the average of all $\bm{m}$, that is the normal vector of the macrosurface, $\bm{n}$, coincided with the $z$-axis with respect to all AFM data and all cutoff wavelength $\lambda_{c}$.
Finally, we calculated the polar angle $\theta_{m}$ of $\bm{m}$ for each of the $256 \times 256$ data points and generated its statistical distribution.
The distribution of $\theta_{m}$ derived from Fig.~\ref{fig5}(b) is illustrated in Fig.~\ref{fig6}.
The value of $\alpha$ was determined by fitting the distribution using Eq.~\ref{equ03}.

\begin{figure}[htbp]
  \centering
  \includegraphics*[width=90mm]{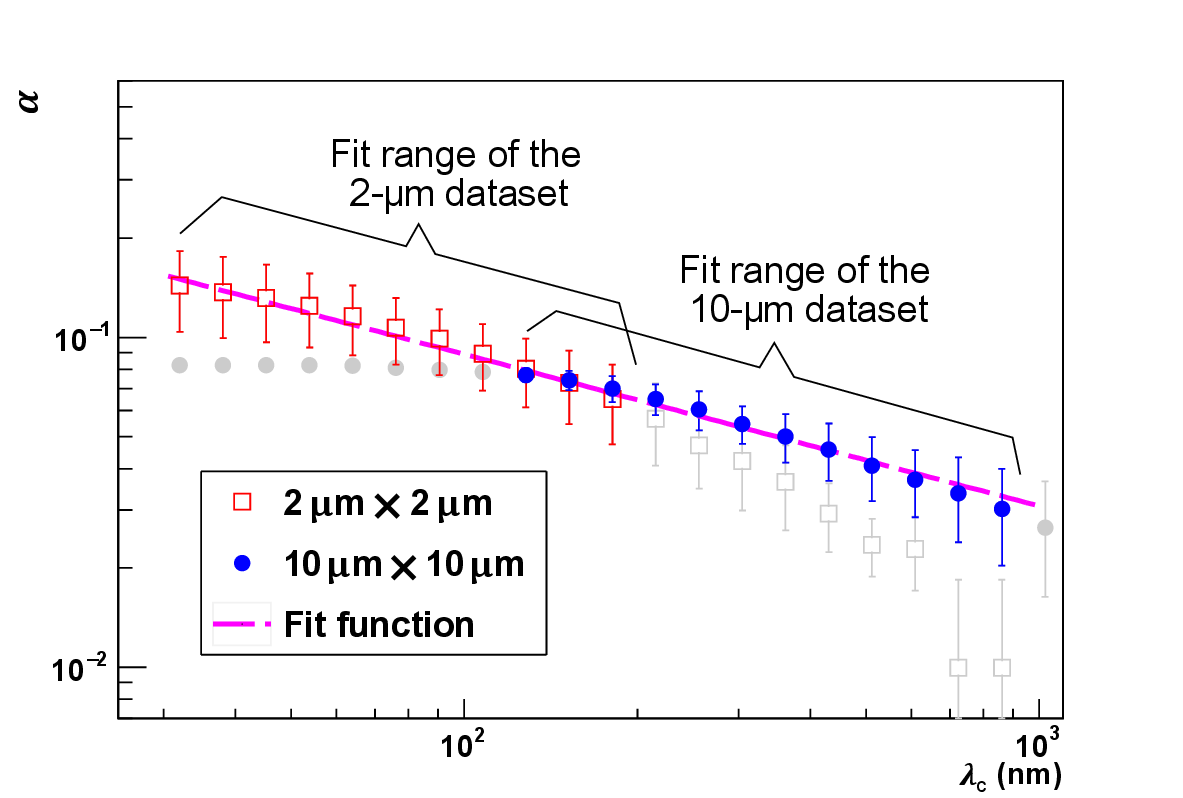}
  \caption{(Color online) Variation of $\alpha$ with respect to $\lambda_{c}$ and the fitted power function.
  The datasets used for the fit are shown in color and by open brackets.\label{fig7}}
\end{figure}

We performed the process for different 21 cutoffs, allowing us to derive the value of $\alpha$ corresponding to each value of $\lambda_{c}$. 
We then applied this to all of the images taken, both at the $2\,\mathrm{\mu m}$ and $10\,\mathrm{\mu m}$ scales, and calculated their averages.
The resulting $\lambda_{c}$ dependence of $\alpha$ is shown in Fig.~\ref{fig7}.
It is worth noting that the data plots show plateaus at short $\lambda_{c}$ due to the limit of spatial resolution of AFM and drops at long $\lambda_{c}$ due to the limit of frequency resolution in the Fourier transform.
We eliminated both ends of the data plots.

As depicted in Fig.~\ref{fig6}, the Beckmann distribution does not perfectly align with the actual $\theta_{m}$ distribution, showing a more rapid decay in the region of larger $\theta_{m}$.
Nevertheless, in this figure case, the average polar angle of the normal vectors, $\langle \theta_{m} \rangle$, is $\langle \theta_{m} \rangle_{\mathrm{data}} = 0.084 \,\mathrm{rad}$ for the actual distribution and $\langle \theta_{m} \rangle_{\mathrm{fit}} = 0.072 \,\mathrm{rad}$ for the fit function, a good agreement, although short by 14\%.
For the data points presented in Fig.~\ref{fig7}, which were provided by averaging several analyses, the shortfall of $\langle \theta_{m} \rangle_{\mathrm{fit}}$ relative to $\langle \theta_{m} \rangle_{\mathrm{data}}$ is at most 17\% and 24\% for the $2\,\mathrm{\mu m}$ and $10\,\mathrm{\mu m}$ scales, respectively.
Thus, the Beckmann distribution provides a good approximation of the actual slope distribution if one allows for the normal distribution to be centered by approximately 20\% relative to reality.

The discrepancy in distribution shape is likely due to the Beckmann model's approach of independently and stochastically determining only the slope of a local surface, without accounting for the continuity of the overall micro-surfaces.
Although this discrepancy could be mitigated by using later models that consider a rough surface as a continuous curved surface~\cite{Walter2007,Trowbridge1975}, we opted to use the Beckmann distribution in our initial step of modeling the large surface waviness.
This decision was made because these improved models are more mathematical and the physical meaning of their roughness parameter is difficult to comprehend.

Referring to~\cite{Mandelbrot1984,Jacobs2017}, the RMS slope, represented by $\alpha$, can be estimated from a power function of the cutoff frequency in Fourier analysis of self-affine surfaces. 
Thus, the data points were fitted with the following power function:
\begin{equation}
  \alpha (\lambda_{c}) = a\lambda_{c}^{\,b}. \label{equ06}
\end{equation}
The fit results are $a=0.74 \pm 0.19$ and $b= -0.46 \pm 0.05$.
The $\alpha(\lambda_{c})$ obtained in this way was used in the simulation.

The mf-BRDF model was implemented in the simulation by loading pre-computed 2D histograms of scattering angle distributions $(\theta_{o}, \phi_{o})$.
A total of 9,000 histograms were created by changing the incident angle $\theta_{i}$ from $0^{\circ}$ to $89^{\circ}$ in $1^{\circ}$ step and the width $\alpha$ from 0 to 0.198 with 0.002 steps.
The angular resolution of the histograms is $1^{\circ}$ for both $\theta_{o}$ and $\phi_{o}$.
At a reflection, the histogram corresponding to $\theta_{i}$ and $\alpha$ was selected, and the scattering angles $(\theta_{o}, \phi_{o})$ were determined using random numbers based on the histogram.

\subsection{\label{sec32} Comparison of experimental results with transport calculations using reflection models}
In the simulation, UCNs were generated so that the velocity distribution of neutrons incident on the guide tube matches with that of the no-guide measurement in Fig.~\ref{fig3}.
The results were normalized to the integrated value of the no-guide measurement.
The coefficient $\eta$ of loss per bounce was set to $2.2 \times 10^{-4}$ based on measurements using the same sample guide~\cite{Schreyer2022}, but since the loss rate due to this effect is $10^{-3}$--$10^{-4}$ in transmittance in our experiments, this contribution is negligible.
In our off-specular scattering algorithm, first, the direction of the reflection vector $\bm{o}$ was determined as described above.
Subsequently, the normal vector $\bm{m}$ was derived from the vector $\bm{o}$ using the formula $\bm{m} = (\bm{o} - \bm{i}) / |\bm{o} - \bm{i}|$ (see Fig.~\ref{fig_ex1}).
Next, it was determined whether a UCN was reflected or transmitted (lost) by the Fermi potential relative to the normal vector $\bm{m}$, or if it was lost due to the loss per bounce effect.
Additionally, the reflected UCN was diffused by the Lambert model with the selection probability $p_{L}$.
Finally, the undiffused UCN was reflected in the direction of the vector $\bm{o}$.
\begin{figure*}[ht]
  \centering
  \includegraphics*[width=170mm]{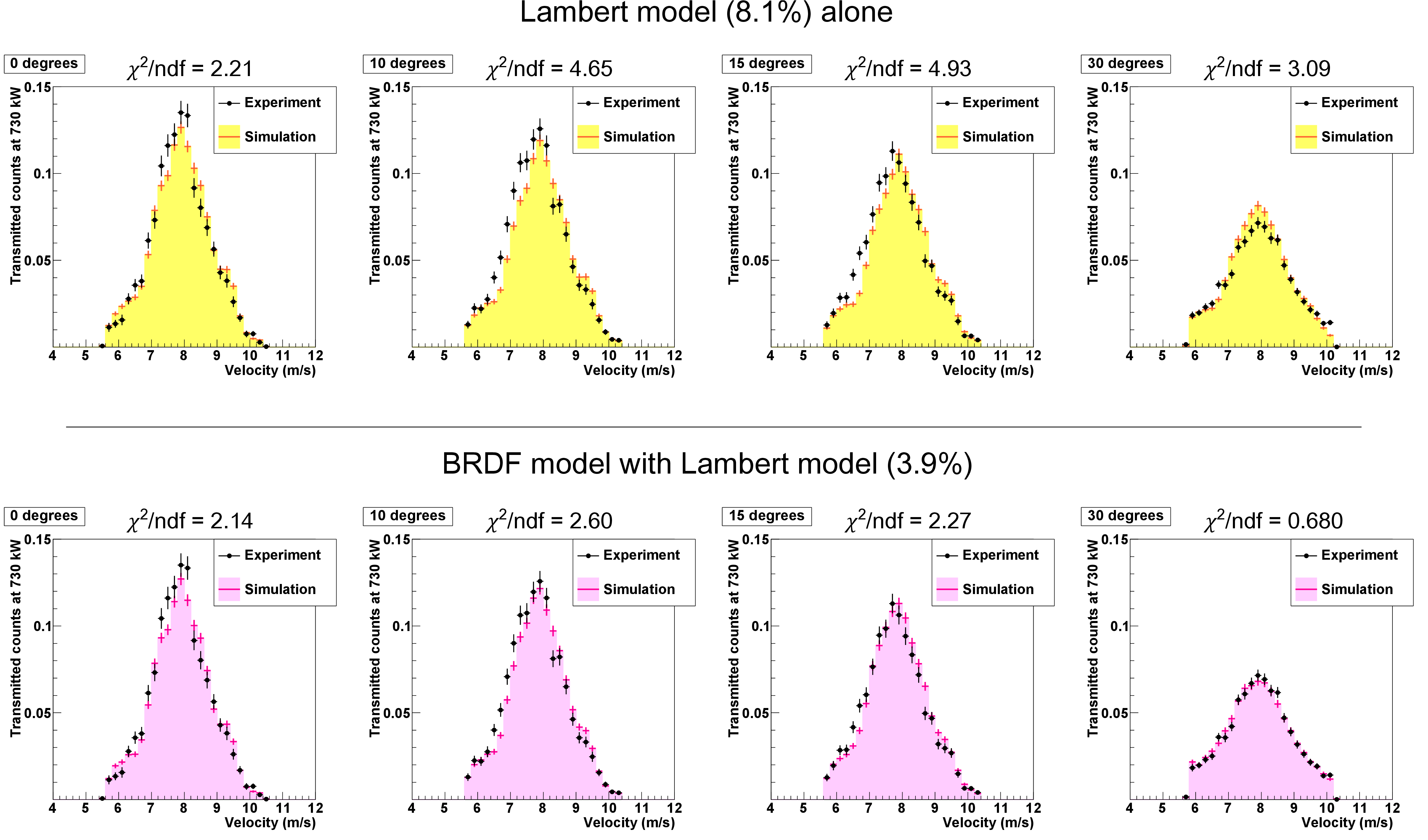}
  \caption{Comparison of velocity distribution shapes between experiment and simulation: simulation with the Lambert model alone with $p_{L} = 0.081$ (top), simulation with the mf-BRDF model combined with the Lambert model with $p_{L} = 0.039$ (bottom). Each figure is accompanied by the $\chi^{2}/\mathrm{ndf}$ values computed from its measured and simulation data sets.\label{fig8}}
\end{figure*}

As shown by the blue triangles in Fig.~\ref{fig4}, the transport efficiency calculated without surface roughness is independent of the installation angle of the guide tube.
The transport efficiency in the no-roughness simulation is reduced compared to the no-guide measurement because of the reduction in diameter between the sample guide tube and detector window.
On the other hand, even at an angle of $0^{\circ}$, this simulated efficiency is still larger than that of the measurement.
The transport efficiencies at a $0^{\circ}$ angle for the two models considering surface roughness described below are in agreement with the experiment.

In performing simulations considering surface roughness, we first attempted to reproduce the measured transport efficiency by using the Lambert model alone.
A least squares fit with $p_{L}$ as a free parameter was performed, and the result is shown in Fig.~\ref{fig4} as orange circles with dotted lines.
The best-fit value for the scattering probability of the Lambert model was $p_{L} = 0.081 \pm 0.003$ and $\chi^{2}/\mathrm{ndf} = 11$.
The transport efficiency decreased monotonically as the installation angle increased, and especially at $10^{\circ}$ and $15^{\circ}$, the transport efficiency deviated significantly from the experimental results.
Therefore, this model could not explain the present experimental results.

A least-squares fit to the measured transport efficiency using a simulation with the mf-BRDF model combined with the Lambert model is shown by the pink squares with dashed lines in Fig.~\ref{fig4}, where $p_{L} = 0.039 \pm 0.003$ was given as the best fit.
The result is in good agreement with the value $p_{L}=0.03$ to 0.05 estimated in our previous work~\cite{Ahmed2019}.
Since $\alpha$ was derived from AFM observations, only $p_{L}$ was considered as a free parameter in this fit.
Its $\chi^{2}/\mathrm{ndf} = 2.1$ means that the simulated transport efficiency and the experimental results are in good agreement.

The TOFs calculated using the best-fit results described above were transformed into velocity distributions using the analytical procedure described in Sec.~\ref{sec2}, and are compared in Fig.~\ref{fig8} with the measured results shown in Fig.~\ref{fig3} (b).
The $\chi^{2}/\mathrm{ndf}$ values obtained from the histograms of the measurement and simulation are also included in each figure.
The integral of the distribution for each simulation corresponds to the data points in Fig.~\ref{fig4}.
The calculation considering the mf-BRDF model reproduces all velocity distributions well with $\chi^{2}/\mathrm{ndf} < 2.6$.
The $\chi^{2}/\mathrm{ndf}$ values were better by 2 than the Lambert model alone for the installation angles larger than $10^{\circ}$.
In particular, the mf-BRDF model reproduces the measured distribution shape very well at an installation angle of $30^{\circ}$.
This is because, unlike the Lambert model in which the specular reflection component remains unchanged, in the mf-BRDF model all specular reflections are diffused lobe-like.
The increase in lobe-like reflections increases the number of UCN incidents on the wall at speeds above the critical velocity of total reflection of $6.3\,\mathrm{m/s}$ for the NiP, resulting in a significant loss increase.
This explains the sharp drop in the transport efficiency at the installation angles of $15^{\circ}$ and $30^{\circ}$.

\section{\label{sec4}Discussions}
\begin{figure}[htbp]
  \centering
  \includegraphics*[width=88mm]{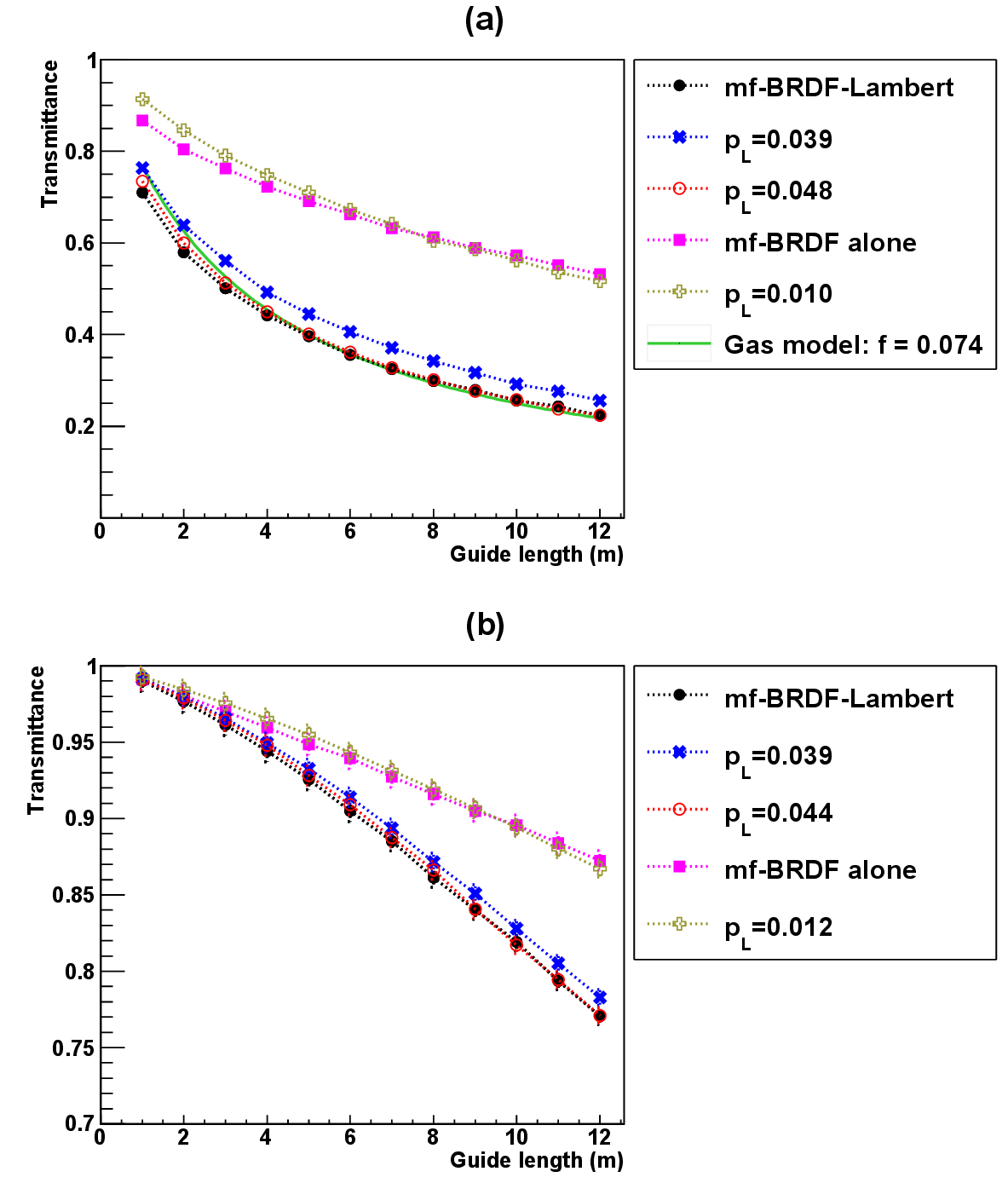}
  \caption{Reduction of the transport efficiency of UCN at a velocity of $5\,\mathrm{m/s}$ with different diffuse scattering models when the length of the guide tube is extended. In (a) the UCN returned to the inlet is assumed to be completely lost (open inlet). The statistical errors are smaller than that of the markers. In (b) the calculation is performed under the condition that the UCN returned to the inlet is reflected by an NiP mirror (closed inlet). See text for calculation settings.
\label{fig9}}
\end{figure}

The mf-BRDF model reproduced the UCN reflection of the guide tube of the TUCAN experiment with very good accuracy.
Therefore, to estimate the UCN transport efficiency in the TUCAN nEDM experiment, we performed transport calculations assuming a straight extension of the length of the sample guide tube to $12\,\mathrm{m}$ in $1\,\mathrm{m}$ increments.
In this calculation, UCN with a velocity of $5\,\mathrm{m/s}$ injected by the cosine distribution law was transported, and realistic joints of the tubes were ignored in order to focus only on the effect of UCN diffusion.
Fig.~\ref{fig9} shows the results of the calculations.
The calculations were performed for the case of complete UCN loss at the entrance of the tube (open inlet: (a) in the figure) and the case of complete reflection (closed inlet: (b) in the figure).
The main cause of UCN loss was the backscattering in the former case and the loss per bounce effect in the latter case.
To evaluate the transport efficiency taking into account the mf-BRDF model, we performed a least squares fit using the Lambert model alone for transport calculations for a range of lengths greater than $6\,\mathrm{m}$.
This estimate resulted in a value of $p_{L}$ that gives the same transport efficiency.

For the open inlet, the decrease in transport efficiency with increasing transport distance using the mf-BRDF combined with the Lambert model with $p_{L}= 0.039$ (the mf-BRDF-Lambert model) is almost consistent with the pure Lambert model with $p_{L}= 0.048$, as shown in Fig.~\ref{fig9} (a).
This $p_{L}$ value is significantly smaller than that of 0.081 derived in Sec.\ref{sec32}.
The reason for this is that the TUCAN guide simulation uses $5\,\mathrm{m/s}$ UCNs, which are totally reflectable, whereas the J-PARC experiment uses faster neutrons that are vertically incident and do not reflect. 
The random surface waviness of the guide tube produces an increase in loss, $\Delta p_{L} = 0.009$, in terms of the Lambert model.
This is also supported by the good agreement between the transport efficiency of the mf-BRDF model alone and the pure Lambert model with $p_{L} = 0.010$, as shown in the same figure.
When the data for the mf-BRDF-Lambert model was fitted with Eq.~\ref{equ01}, it yielded $f=0.074$, which is about twice the value of the scattering probability set for the Lambert model in the simulation.

For the closed inlet shown in (b), the transmission attenuation for the mf-BRDF model alone and the pure Lambert model with $p_{L} = 0.012$ are almost identical, while the value of $p_{L}$ consistent with the mf-BRDF-Lambert model is 0.044.
Hence, in this case, the surface waviness of the sample guide tube causes only $\Delta p_{L} = 0.005$.
The mf-BRDF model induces a slight deviation in the direction of scattered UCNs from the specular direction, whereas the Lambert model scatters UCNs in all directions.
As a result, when UCNs are scattered by both models for a long time, the directional motion unique to the mf-BRDF model is expected to gradually fade away, and the UCN motion direction is anticipated to become completely diffusive.
In the case where the UCN inlet is closed by an NiP mirror, UCNs remain in the guide tube for a longer period compared to the open inlet because backscattered UCNs are reflected at the inlet.
Consequently, in these calculations, the direction of UCN motion would be more randomized, and the contribution of the mf-BRDF model would be weakened.

Recent UCN sources are designed to store UCNs, but inevitably, there is UCN loss due to up-scattering or nuclear absorption~\cite{Ahmed2019,Bison2022}.
Consequently, they can generally be considered to be between the open and closed inlet cases. 
In any case, it is assumed that the UCN loss in the guide tubes, produced by the combination of surface roughness from polishing by Irving Polishing \& Manufacturing Inc. and coating by Chem Processing Inc. which we are planning to use for the TUCAN experiment, is dominated by Lambertian diffusion caused by surface microstructure rather than surface waviness.
Therefore, the removal of microstructure is a priority issue.
However, since surface waviness alone also causes a reduction in transport efficiency as large as in the Lambert model with $p_{L} = 0.01$, it must eventually be eliminated as well.

Our model uses $\alpha$ values determined from AFM measurements and $p_{L}$ values obtained from experiments.
The scattering probability of the Lambert model, $p_{L}$, is a free parameter and is not determined from the images observed by the AFM.
This is because the roughness of the guide tubes used in the experiment is large with an RMS amplitude of $b = 6.4$--$17\,\mathrm{nm}$, and there is no scattering model that can directly describe this surface at any scale~\cite{Steyerl1972, Brown1975}.
For example, if the RMS amplitude is about $b = 1\,\mathrm{nm}$, it can be described by the MR model as reported by~\cite{Atchison2010}.
In the future, the description of UCN scattering by structures smaller than the incident wavelength, which was ignored in the present modeling, could be achieved by analytical calculations using the DWBA~\cite{Messiah1958, Sinha1988, Pynn1992} or by a direct numerical calculation of wave equations relying on the processing power of a computer.
By combining the results with the mf-BRDF model, as in a two-scale reflectance model of light~\cite{Holzschuch2017}, it could be possible to describe the UCN scattering including the effect of waviness from an appropriate AFM image.

\section{\label{sec5}Conclusions}
We have performed transport experiments of pulsed UCN beams at BL05 in J-PARC/MLF to search for a model that can accurately describe the UCN transport for a typical guide tube, including the off-specular scattering. 
The measured guide tube was fabricated by an aluminum structure with NiP plating as used in many UCN experiments.
In the experiments, pulsed UCN beams with divergence angles of $\pm 6^{\circ}$ or less were injected into a UCN guide tube with an internal diameter of $95.5\,\mathrm{mm}$ and a length of $1000\,\mathrm{mm}$ installed at angles of $0^{\circ}$, $10^{\circ}$, $15^{\circ}$, and $30^{\circ}$ relative to the incident direction.
The decrease in transport efficiency and the deformation of the TOF spectrum with increasing angles were measured.
A least-squares fit to the measured UCN transport efficiencies was performed using simulations implementing the Lambertian diffusion with the diffusion probability $p_{L}$ as a free parameter, and the result was $p_{L} = 0.081\pm0.003$ at $\chi^{2}/\mathrm{ndf} = 11$.
For the degree of consistency between the shape of the measured and simulated TOF spectra, the results were $\chi^{2}/\mathrm{ndf} = 3.1$--4.9 at angles larger than $10^{\circ}$.
Therefore the measured results were not explained by the Lambert model.

We developed a new UCN scattering model based on AFM measurements to explain the experimental results.
In our model, surface waviness larger than the UCN wavelength is extracted by Fourier analysis, and the distribution of the surface normals is incorporated into a mf-BRDF model based on geometrical optics.
The scattering from the surface waviness is described by the mf-BRDF model, and the scattering from smaller structures and other factors is described by the Lambert model. 
$p_{L}$ was set as a free parameter again, and fitting the model simulation to the measured transport efficiency yielded $p_{L} = 0.039\pm0.003$ at $\chi^{2}/\mathrm{ndf} = 2.1$.
The TOF shapes calculated using this value were consistent with the measured TOF with $\chi^{2}/\mathrm{ndf} < 2.6$, and the $\chi^{2}/\mathrm{ndf}$ values were better by 2 than the Lambert model alone for installation angles larger than $10^{\circ}$.
Therefore, we have succeeded in finding a model that reproduces the transport efficiency of NiP-plated aluminum UCN guide tubes and the off-specular scattering on their surfaces well.

In the transport calculations with a $12\,\mathrm{m}$ long guide tube and an incident UCN velocity of $5\,\mathrm{m/s}$, the reduction in transport efficiency due to the surface waviness was estimated to be comparable to the Lambert model with $p_{L} \sim 0.01 $.
Therefore, polishing the smaller structures rather than the surface waviness was found to be the priority issue for this guide tube.

\begin{acknowledgments}
This research was supported by JSPS KAKENHI Grant Number 18H05230, 20KK0069, and 21K13940. The neutron experiment at the Materials and Life Science Experimental Facility of the J-PARC was performed under user programs (Proposal No. 
2021B0272 and 2021B0309) and S-type project of KEK (Proposal No. 2019S03).
\end{acknowledgments}

\bibliography{references}

\end{document}